\documentclass[aps, prb, twocolumn]{revtex4}
\usepackage{graphics}
\usepackage{graphicx}
\usepackage{color}
 
\draft

\begin{document}

\title{Carbon Nanotube with Square Cross-section: An \textit{Ab Initio} Investigation}
 
\author{P A S Autreto}
\email{autretos@ifi.unicamp.br}
\affiliation{Instituto de F\'{i}sica ``Gleb Wataghin'', Universidade Estadual de Campinas, Unicamp, C.P. 6165, 13083-970, Campinas, S\~{a}o Paulo, Brazil}
 
\author{S B Legoas}
\affiliation{Centro de Ci\^{e}ncias e Tecnologia, Universidade Federal de Roraima, 69304-000, Boa Vista, Roraima, Brazil}
 
\author{M Z S Flores}
\affiliation{Instituto de F\'{i}sica ``Gleb Wataghin'', Universidade Estadual de Campinas, Unicamp, C.P. 6165, 13083-970, Campinas, S\~{a}o Paulo, Brazil}
 
\author{D S Galvao}
\email{galvao@ifi.unicamp.br}
\affiliation{Instituto de F\'{i}sica ``Gleb Wataghin'', Universidade Estadual de Campinas, Unicamp, C.P. 6165, 13083-970, Campinas, S\~{a}o Paulo, Brazil}

\date{\today}

\begin{abstract}
Recently, Lagos et al. (Nature Nanotechnology $\textbf{4}$, 149 (2009)) reported the discovery of the smallest possible silver square cross-section nanotube.
A natural question is whether similar carbon nanotubes can exist. In this work we report \textit{ab initio} results for the structural, stability and electronic properties for such hypothetical structures. Our results show that stable (or at least metastable) structures are possible with metallic properties. They also show that these structures can be obtained by a direct interconversion from SWNT(2,2). Large finite cubane-like oligomers, topologically related to these new tubes were also investigated.
\end{abstract}
 
\maketitle
 
\section{Introduction}
 
The study of the mechanical properties of nanoscale systems presents new theoretical and experimental challenges \cite{ray,book}. The arrangements of atoms at nano and macro scales can be quite different and affect electronic and mechanical properties. Of particular interest are the structures that do not exist at macroscale but can be formed (at least as metastable states) at nanoscale, specially when significant stress/strain is present. Examples of these cases are atomic suspended chains \cite{onishi,ruit,alloy} and helical nanowires \cite{tosatti}.
 
\begin{figure}[h]
\begin{center}
\includegraphics[scale=0.4]{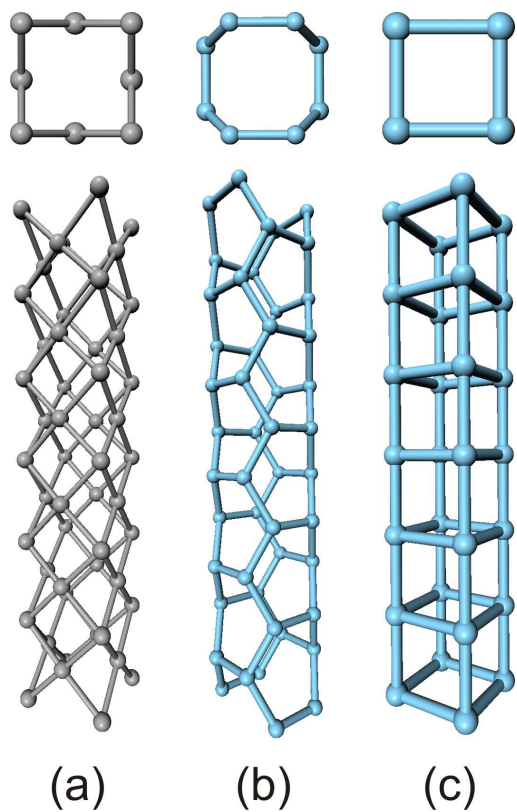}
\caption{(color online). Structural nanotube models. Frontal and lateral views: (a) Silver tube \cite{agtube}, (b) SWNT(2,2), and (c) CNTSC. See text for discussions.}
\label{figure1}
\end{center}
\end{figure}

More recently \cite{agtube}, it was discovered the smallest metal (silver) nanotube  from high resolution transmission electron microscopy (HRTEM) experiments. These tubes are spontaneously formed during the elongation of silver nanocontacts. \textit{Ab initio} theoretical modeling \cite{agtube} suggested that the formation of these hollow structures requires a combination of minimum size and high gradient stress. This might explain why these structures had not been discovered before in spite of many years of investigations. Even from theoretical point of view, where low stress regimes and small structures have been the usual approach, no study predicted their existence. The unexpected discovery of this new family of nanotubes suggests that such other `exotic' nanostructures may exist. One question that naturally arises is whether carbon-based similar nanotubes (i.e., carbon nanotubes with square cross-section - CNTSC) could exist (figure \ref{figure1}).

\begin{figure}[h]
\begin{center}
\includegraphics[scale=0.3]{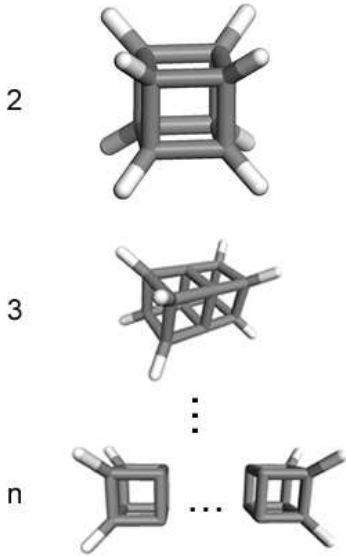}
\caption{(color online). Cubane and its `polymerized' units. The label refers to the number of square cross-sections in the structure. We considered structures from cubane up to n=10. Stick models, C and H atoms are in grey and white colors, respectively.}
\label{figure2}
\end{center}
\end{figure}

From the topological point of view CNTSC tubes would require carbon atoms arranged in multiple square-like configurations. Molecular motifs satisfying these conditions, the so-called cubanes, do exist and they are stable at room temperature (figure \ref{figure2})  \cite{cub1}.   Cubane (C$_8$H$_8$) is a hydrocarbon molecule consisting of 8 carbon atoms in an almost perfect cubic arrangement (thus its name). Hydrogen atoms are bonded to each carbon atom (figure \ref{figure2}), completing its four-bond valency. The 90 degrees angles among carbon atoms create a very high strained structure. Because of this unusual carbon hybridization, cubane was considered a `platonic' hydrocarbon and believed to be almost impossible to be synthesized \cite{cub1}. However, in 1964 Eaton and Cole \cite{cub2} achieved its breakthrough synthesis. Since then the cubane chemistry evolved quickly \cite{cub1,cub3}. Solid cubane \cite{cub4} proved to be remarkably stable and polymers containing up to substituted 40 cubanes units have been already synthesized \cite{cub1}. However, up to now no tubular structure has been reported \cite{cub5,cub6}.  

In this work we have theoretically investigated structural, stability and electronic properties of CNTSC tubes. We have considered infinite (considering periodic boundary conditions) and finite (oligomers up to ten square units, figure \ref{figure2}) structures.
 
\section{Methodology}
 
We have carried out \textit{ab initio} total energy calculations in the framework of density functional theory (DFT), as implemented in the DMol$^3$ code \cite{dmol3}. Exchange and correlation terms were treated within the generalized gradient (GGA) functional by Perdew, Burke, and Ernzerhof \cite{pbe}. Core electrons were treated in a non-relativistic all electron implementation of the potential. A double numerical quality  basis set  with polarization function (DNP) was considered, with a real space cutoff of 3.7 \AA. The tolerances of energy, gradient, and displacement convergence were 0.00027 eV, 0.054 eV/\AA\, and 0.005 \AA, respectively.
Initially we optimized the CNTSC unit cell with fixed $\textit{a}$ and $\textit{b}$ parameters set to 20 \AA\, in order to ensure isolated (non-interacting) structures. The axial $\textit{c}$ lattice parameter was varied, and the total energy per unit cell calculated. All internal atomic coordinates were free to vary. Total energy versus unit cell volume was fitted using the Murnaghan procedure \cite{murnaghan} to obtain the equilibrium \textit{c} lattice parameter. For comparison purposes we have also considered graphite, diamond and carbon nanotube SWNT(2,2). SWNT(2,2) was chosen because, although ultra-small carbon nanotubes (USCNTs) have been theoretically investigated \cite{cox,scipioni,yuan,kuzmin}, it remains the smallest CNT experimentally observed with an estimated diameter of 3 \AA\ \cite{22prl1,22prl2}. Graphite and diamond were also included in our study, because they are the two most stable carbon form and to provide a benchmark for the relative stability between the different tubes and these structures.

\section{Results and Discussions}
 
The results are presented in table \ref{table1}. As expected, graphite is the structure with the lowest energy (most stable), followed by diamond, SWNT(2,2) and CNTSC, respectively. Although the CNTSC energy per atom is relatively high (in part due to the strained C-C bonds, as in cubanes) its relative energy difference to SWNT(2,2) (0.0384 Ha) is similar to the difference between SWNT(2,2) and cubic diamond (0.0395 Ha).

\begin{table}[t]
\centering
\caption{\label{table1}DMol$^3$ results for crystalline carbon allotropic structures:
(a) Graphite, (b) Cubic diamond, (c) SWNT(2,2), and (d) CNTSC.}
\begin{tabular}{|ccccc|}
\hline
    & (a) & {(b)} & (c) & (d) \\ 
    \hline
Energy/atom (Ha)  & $-38.085$ & $-38.081$ & $-38.041$ & $-38.003$ \\ 
Lattice parameters: & & & &\\
\textit{a} (\AA)  & $2.46$ & $3.57$ & $20.0$ & $20.0$ \\ 
\textit{c} (\AA)  & $6.80$ & $3.57$ & $2.53$ & $1.62$ \\ 
C-C bond-length (\AA): & $1.423$ & $1.537$ & $1.447$ & $1.580$ \\
  &  &  & $1.464$ & $1.616$ \\
  \hline
\end{tabular}
\end{table}

\begin{figure}[b]
\begin{center}
\includegraphics[scale=0.6]{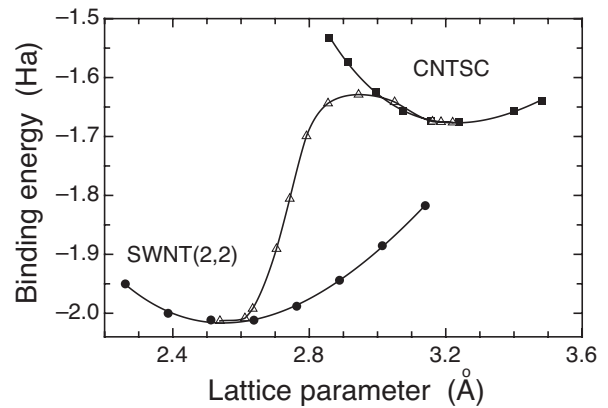}
\caption{Binding energy per unit cell as a function of axial \textit{c} lattice parameter for SWNT(2,2) (circles) and CNTSC (squares). It is also shown its interconversion curve (triangles). See text for discussions.}
\label{figure3}
\end{center}
\end{figure}

In figure \ref{figure3} we present the binding energy per unit cell. As a direct comparison it is not possible because the number of atoms in the minimum unit cell is different for SWNT(2,2) and CNTSC (eight and four, respectively), we used a double CNTSC unit cell. As can be seen from figure \ref{figure3}, the results suggest that stable (or at least metastable, as a well defined minimum is present) CNTSC structures are possible.

Our results also suggest that a direct interconversion from SWNT(2,2) to CNTSC is unlikely to occur via axial(longitudinal) stretching. The extrapolation of the stretched SWNT(2,2) curve (figure \ref{figure3}, circle data points) could be misleading suggesting that it would be possible an interception with the stretched CNTSC curve (figure \ref{figure3}, square data points). However, this did not occur, the SWNT(2,2) can not preserve its tubular topology when its c-value is beyond 3.2 \AA. 

\begin{figure}[t]
\begin{center}
\includegraphics[scale=0.4]{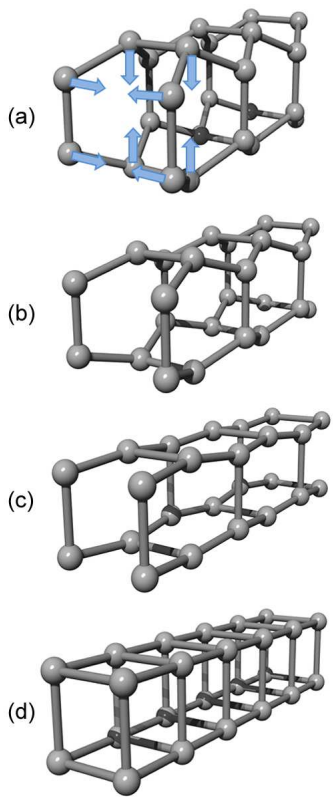}
\caption{(color online). Snapshots from the axial compression process, showing the interconversion of SWNT(2,2) to CNTSC. (a) Initial SWNT(2,2), (b) and (c) intermediates, and (c) final CNTSC structures.}
\label{figure4}
\end{center}
\end{figure}

We then investigated whether if an assisted interconversion would be possible, in our case we considered a continuously decrease of the tube radii value (in order to mimic an applied external (radially)   pressure) while keeping the tube free to expand/contract longitudinally (see figure \ref{figure4} and video1 \cite{EPAPS}).

We have performed these calculations starting from an optimized SWNT(2,2) unit cell and then continuously decreasing its radii value and re-equilibrating the system and measuring the new c-values (figure \ref{figure3}, triangle data points). Our results show that under these conditions there is a pathway that permits a direct interconversion from SWNT(2,2) to CNTSC.
  
In figure \ref{figure4} we present a sequence of snapshots from the simulations representing the interconversion process. The strain energy injected into the system by the radial compression (figure \ref{figure4}(a))  produces a $\textit{c}$-lattice expansion, leading to a structural transition (figure \ref{figure4}(d)). The compression 
process produces new C-C bonds followed by carbon rehybridizations.
The processes is better visualized in the supplementary materials (video1) \cite{EPAPS}.

\begin{figure}[b]
\begin{center}
\includegraphics[width=6cm]{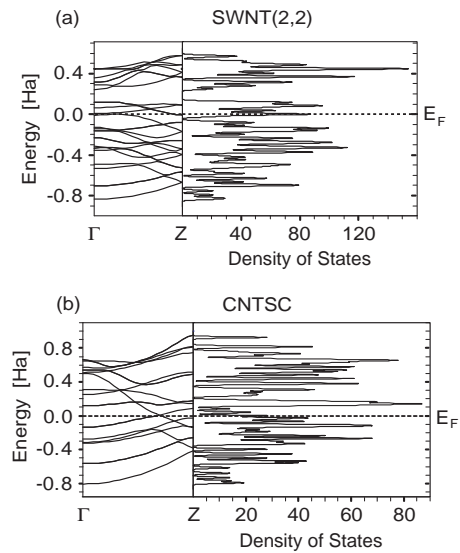}
\caption{Band structure and total density of states (in electrons/Ha) results for
the (a) SWNT(2,2) and (b) carbon square-cross-section CNTSC. 
Energy is relative to Fermi level (dashed horizontal lines). Primitive unit cells have eight and four carbon atoms for SWNT(2,2) and CNTSC,
respectively.}
\label{figure5}
\end{center}
\end{figure}

We then proceeded with the CNTSC electronic structure analysis. In figure \ref{figure5} we displayed the band structure and the density of states (DOS) for SWNT(2,2) and CNTSC tubes. Both structures present a finite DOS at the Fermi energy, characteristic of metallic regimes. Although CNTSC exhibits non-usual carbon hybridization it follows the general trends that small diameter carbon nanotubes are metallic \cite{metallic,kamal}.

It is possible that synthetic methods used to produce CNTs (such as laser ablation, chemical vapor deposition and arc discharge \cite{book}) could also produce CNTSC, specially inside CNTs of different chiralities, as in the case of SWNT(2,2) \cite{22prl1,22prl2}. Another possibility could be a polymeric synthetic approach, like the topochemical ones to produce carbon nanotube of specific types that have been recently discussed in the literature \cite{raytopo,jacstopo}. Considering that cubane molecules and their polymers exist and are stable, we decided to investigate the local stability and endothermicity of cubane-like tubular oligomers that are topologically related to CNTSC.  We carried out DMol$^3$ calculations for the molecular structures displayed in figure \ref{figure2}. The terminal C atoms were passivated with H atoms.

\begin{table}[t]
\centering
\caption{\label{table2}Total energy per carbon atom (in Ha) for the structures shown in figure \ref{figure2}. The corresponding value for the infinite structure is also presented.}
\begin{tabular}{|c|c|}
\hline
  $\textit{n}$ &  $e_t(n)$ \\ 
\hline
{2} &-38.647 \\
\hline 
{3} & {-38.429} \\
\hline
{4} & {-38.323} \\ 
\hline
{5} & {-38.259} \\
\hline 
{6} & {-38.216} \\
\hline
{7} & {-38.186} \\
\hline
{8} & {-38.163} \\
\hline
{9} & {-38.145} \\
\hline
{10} & {-38.131} \\
 
{$\vdots$} & {$\vdots$} \\
{CNTSC} & {-38.003} \\
\hline
\end{tabular}
\end{table}
 
In table \ref{table2} we present the results for the total energy per carbon atom as a function of the number of square cross-sections. Our results show that although the oligomer formation would require an endothermal (energetically assisted) process, the structures are stable and the energy per carbon atom converges assyntotically to the corresponding value of the infinite tube (see also supplementary materials \cite{EPAPS}).

In summary, based on a recent discovery of the smallest possible silver nanotube with a square cross-section \cite{agtube}, we have investigated whether a similar carbon-based structure could exist. We have used \textit{ab initio} methodology to investigate the structural, stability and electronic properties of carbon nanotubes with square cross-section (CNTSC). Our results show that stable (or at least metastable) CNTSC (finite and infinite) structures can exist. They also show that it is possible to convert  SWNT(2,2) to CNTSC under radial contraction. CNTSCs should share most of the general features exhibited by 'standard' nanotubes. Although the CNTSCs have not yet been observed, we believe our results had proven their feasibility.We hope the present work will estimulate further works for these new family of carbon nanotubes.

\begin{acknowledgments}
 
This work was supported in part by the Brazilian Agencies CNPq, CAPES and FAPESP. DSG thanks Prof. D. Ugarte for helpful discussions.
 
\end{acknowledgments}
 

\end{document}